\documentclass[12pt,fleqn]{article} 
\usepackage{graphicx,color,amsfonts}
\newcommand{\be}{\begin{equation}}
\newcommand{\ee}{\end{equation}}
\newcommand{\bea}{\begin{eqnarray}}
\newcommand{\eea}{\end{eqnarray}}

\newcommand{\ep}{\qquad {\vrule height 10pt width 8pt depth 0pt}}

\newcommand{\grintl}{[\kern-.18em [}
\newcommand{\grintr}{]\kern-.18em ]}
\newcommand{\ds}{\displaystyle}
\newtheorem{theorem}{Theorem}[section]
\newtheorem{thm}[theorem]{Theorem}

\newtheorem{lem}[theorem]{Lemma}

\def\smallR{\hbox{\scriptsize I\kern-.23em{R}}}
\def\R{\hbox{$\mit I$\kern-.33em$\mit R$}}
\def\C{\hbox{$\mit I$\kern-.6em$\mit C$}}
\def\un{\hbox{$\mit I$\kern-.77em$\mit I$}}
\def\0{\hbox{$\mit I$\kern-.70em$\mit O$}}
\def\r{I\kern-.277em R}

\begin{document}

\title{Generating Function and a Rodrigues Formula
for the Polynomials in $d$--Dimensional Semiclassical Wave Packets}

\author{George A. Hagedorn\thanks{Partially
Supported by National Science Foundation
Grant DMS--1210982.}\\
Department of Mathematics and\\
Center for Statistical Mechanics and Mathematical Physics\\
Virginia Polytechnic Institute and State University\\
Blacksburg, Virginia 24061-0123, U.S.A.\\ \\
E--mail:\quad hagedorn@math.vt.edu}

\maketitle

\vskip 1cm
\begin{abstract}
We present a simple formula for the generating function
for the polynomials in the $d$--dimensional semiclassical wave packets.
We then use this formula to prove the associated Rodrigues formula.
\end{abstract}

\baselineskip=20pt 
\parskip=9pt

\vskip 1cm
\section{Introduction}
\setcounter{equation}{0}
The generating function for 1--dimensional semiclassical wave
packets is presented in formula (2.47) of \cite{raise}.
In this paper, we present and prove the $d$--dimensional analog.
As an application of this formula,
we also prove the associated (multi--dimensional) Rodrigues formula.

These results have also been proven from a completely different point of view
by Helge Dietert, Johannes Keller, and
Stephanie Troppmann. See Lemma 3 and Section 3 (particularly
Proposition 16) and formula (13) of \cite{DKT}. See also \cite{Keller}.
We have also received a conjecture from
Tomoki Ohsawa \cite{O} that the generating function
result could be proved abstractly by
using the formula for products of Hermite polynomials
and the action of the metaplectic group.

\vskip 3mm
\noindent
{\large\bf Acknowledgements}\quad It is a pleasure to thank Raoul Bourquin
and Vasile Gradinaru for motivating this work.  It is also a pleasure to thank
Johannes Keller, Tomoki Ohsawa, Sam Robinson, and Leonardo Mihalcea
for their enthusiasm and numerous comments.  It is also a pleasure to thank
the referee whose comments improved this paper significantly.
The author was supported in part by National Science Foundation grant
DMS--1210982.

\section{Semiclassical Wave Packets}
The semiclassical wave packets depend on two invertible
$d\times d$ complex matrices $A$ and $B$ that are
always assumed to satisfy
$$
A^*\,B+B^*\,A\ =\ 2\,I\qquad\mbox{and}\qquad A^t\,B-B^t\,A\ =\ 0.
$$
They also depend on a phase space point $(a,\,\eta)$ that plays no role in
the present work.
After chosing a branch of the square root, we define
\bea\nonumber
\varphi_0(A,\,B,\,\hbar,\,a,\,\eta,\,x)&=&
\pi^{-d/4}\,\hbar^{-d/4}\,(\det\,A)^{-1/2}
\\[3mm]\nonumber
&&\hspace{5mm}\times\quad
\exp\left(-\,\frac{\langle(x-a),\,B\,A^{-1}\,(x-a)\rangle}{2\,\hbar}\,+\,i\,
\frac{\langle\eta,\,(x-a)\rangle}{\hbar}\right).
\eea
Here, and for the rest of this paper,
we regard ${\mathbb R}^d$ as being embedded in ${\mathbb C}^d$, and
for any two vectors $a\in{\mathbb C}^d$ and $b\in{\mathbb C}^d$,
we use the notation
$$
\langle\,a,\,b\,\rangle\ =\ \sum_{j=1}^d\
\overline{a_j}\ b_j.
$$

For $1\le l\le d$, we define the $l^{\mbox{\scriptsize th}}$ raising operator
$$
{\cal R}_l\ =\ 
{\cal A}_l(A,\,B,\,\hbar,\,a,\,\eta)^*\ =\
\frac 1{\sqrt{2\,\hbar}}\,\left(
\langle\,B\,e_l,\,(x-a)\,\rangle\,-\,i\,
\langle\,A\,e_l,\,\left(-\,i\,\hbar\nabla-\eta\right)\,\rangle\right).
$$
Then recursively, for any multi-index $k$, we define
$$
\varphi_{k+e_l}(A,\,B,\,\hbar,\,a,\,\eta,\,x)\ =\
\frac 1{\sqrt{k_l+1}}\ {\cal R}_l(\varphi_k(A,\,B,\,\hbar,\,a,\,\eta))(x).
$$

For fixed $A,\,B,\,\hbar,\,a,\,\eta$, these wave packets form
an orthonormal basis indexed by $k$. It is easy to see that
$$
\varphi_k(A,\,B,\,\hbar,\,a,\,\eta,\,x)\ =\
2^{-|k|/2}\,(k!)^{-1/2}\,P_k(A,\,\hbar,\,(x-a))\,
\varphi_0(A,\,B,\,\hbar,\,a,\,\eta,\,x),
$$
where $P_k(A,\,\hbar,\,(x-a))$ is a polynomial of degree $|k|$ in
$(x-a)$, although from this definition, it is not immediately
obvious that $P_k(A,\,\hbar,\,(x-a))$ is independent of $B$.

Since they play no interesting role in what we are doing here,
we henceforth assume $a=0$ and $\eta=0$.

\section{The Generating Function}
Our main result for the generating function is the following:
\begin{thm}
The generating function for the family of
polynomials~ $P_k(A,\,\hbar,\,x)$~ is
$$
G(x,\,z)\ =\
\exp\left(\,-\,\left\langle\,\overline{z},\,A^{-1}\,\overline{A}\,z\right\rangle
\,+\,\frac 2{\sqrt{\hbar}}\,
\left\langle\,\overline{z},\,A^{-1}\,x\right\rangle\right).
$$
{\em I.e.},
$$
G(x,\,z)\ =\ \sum_k\ P_k(A,\,\hbar,\,x)\ \frac{z^k}{k!}.
$$
\end{thm}

\vskip 3mm
\noindent{\bf Remark}\quad We make the unconventional
definition $|A|=\sqrt{A\,A^*}$. By our conditions on the matrices
$A$ and $B$, this forces $|A|$ to be real symmetric and strictly positive.
We also have the polar decomposition $A=|A|\,U_A$,
where $U_A$ is unitary.
With this notation, we can write
$$
G(x,\,z)\ =\
\exp\left(\,-\,\left\langle\,U_A\,\overline{z},\,\overline{U_A}\,z\right\rangle
\,+\,\frac 2{\sqrt{\hbar}}\,\left\langle\,U_A\,\overline{z},\,
|A|^{-1}\,x\right\rangle\right).
$$
This equivalent formula is the one we shall actually prove.

We begin the proof of Theorem 3.1
with a lemma that provides an alternative formula for ${\cal R}_l$.
From this formula and an induction on $|k|$, one can easily prove that
$P_k(A,\,\hbar,\,x)$ is independent of $B$, because
$$
\overline{\varphi_0(A,\,B,\,\hbar,\,0,\,0,\,x)}\
\varphi_0(A,\,B,\,\hbar,\,0,\,0,\,x)\ =\
\pi^{-d/2}\,\hbar^{-d/2}\,|\det A|^{-1}\,
\exp\left(-\,\frac{\langle x,\,|A|^{-2}\,x\rangle}\hbar\right).
$$

\begin{lem}For any $\psi\in{\cal S}$,
$$
(R_l\,\psi)(x)\ =\ -\ \sqrt{\frac\hbar 2}\
\frac 1{\,\overline{\varphi_0(A,\,B,\,\hbar,\,0,\,0,\,x)}\,}\
\left\langle\,A\,e_l,\, \nabla\,
(\,\overline{\varphi_0(A,\,B,\,\hbar,\,0,\,0,\,x)}\ \psi(x))\,\right\rangle.
$$
\end{lem}

\noindent{\bf Proof}:\quad The gradient on the right hand side
of the equation in the lemma can act either on the $\overline{\varphi_0}$ or
on the $\psi$.  So, we get two terms when we compute this:
\begin{eqnarray}\nonumber
&&\sqrt{\frac{\hbar}2}\ \Bigg(\,
\frac 1{2\,\hbar}\,\sum_{j=1}^d\,
\left\langle\,A\,e_l,\
\left(e_j\,\left(\langle\,e_j,\,\overline{B}\,\overline{A}^{\ -1}\,
x\rangle\,+\,
\langle\,x,\,\overline{B}\,\overline{A}^{\ -1}\,
e_j\rangle\right)\right.\right\rangle\,\psi(x)
\\[3mm]\nonumber
&&\hspace{10cm}-\quad
\left\langle\,A\,e_l,\ (\nabla\,\psi)(x)\,\right\rangle\Bigg).
\end{eqnarray}
The second term here is precisely the second term
$\displaystyle \frac 1{\sqrt{2\,\hbar}}\,\left(-\,i\,
\langle\,A\,e_l,\,\left(-\,i\,\hbar\nabla\right)\psi(x)\,\rangle\right)$,
in the expression for $(R_l\psi)(x)$.
So, we need only show the first term here equals the first term,
$\displaystyle \frac 1{\sqrt{2\,\hbar}}\,\langle\,B\,e_l,\,x\,\rangle\,\psi(x)$,~
in the expression for $(R_l\psi)(x)$.

To do this, we begin by noting that the first term here equals
\begin{eqnarray}\nonumber
&&\frac 1{2\,\sqrt{2\,\hbar}}\ \sum_{j=1}^d\
\left\langle\,A\,e_l,\
\left(e_j\,\left(\langle\,e_j,\,\overline{B}\,\overline{A}^{\ -1}\,
x\rangle\,+\,
\langle\,x,\,\overline{B}\,\overline{A}^{\ -1}\,e_j\rangle\right)
\right.\right\rangle\,\psi(x)
\\[3mm]\nonumber
&=&
\frac 1{2\,\sqrt{2\,\hbar}}\ \sum_{j=1}^d\
\left\langle\,A\,e_l,\
\left(e_j\,\left(\langle\,e_j,\,\overline{B}\,\overline{A}^{\ -1}\,
x\rangle\,+\,
\overline{\langle\,\overline{B}\,\overline{A}^{\ -1}\,
e_j,\,x\,\rangle}\right)\right.\right\rangle\,\psi(x)
\\[3mm]\nonumber
&=&
\frac 1{2\,\sqrt{2\,\hbar}}\ \sum_{j=1}^d\
\left\langle\,A\,e_l,\
\left(e_j\,\left(\langle\,e_j,\,\overline{B}\,\overline{A}^{\ -1}\,
x\rangle\,+\,
\langle\,B\,A^{-1}\,
e_j,\,x\,\rangle\right)\right.\right\rangle\,\psi(x)
\\[3mm]\nonumber
&=&
\frac 1{2\,\sqrt{2\,\hbar}}\ \sum_{j=1}^d\
\left\langle\,A\,e_l,\
\left(e_j\,\left(\langle\,e_j,\,\overline{B}\,\overline{A}^{\ -1}\,
x\rangle\,+\,
\langle\,
e_j,\,\left(A^{-1}\right)^*\,B^*\,x\,\rangle\right)\right.\right\rangle\,\psi(x)
\\[3mm]\nonumber
&=&
\frac 1{\sqrt{2\,\hbar}}\,
\left\langle\,A\,e_l,\
\frac{\overline{B}\,\overline{A}^{\ -1}\,+\,
\left(A^{-1}\right)^*\,B^*}2\ x\,\right\rangle
\,\psi(x)
\end{eqnarray}
Because of the relations satisfied by $A$ and $B$,~\, $B\,A^{-1}$ is
(real symmetric)$\,+\,i\,$(real symmetric).
So, its conjugate,~ $\overline{B}\ \overline{A}^{\ -1}$~ has this same form.
Thus,~ $\overline{B}\ \overline{A}^{\ -1}$~ equals its transpose, which is~
$\left(A^{-1}\right)^*\,B^*$.~
So, the quantity of interest here equals
\begin{eqnarray}\nonumber
&&\frac 1{\sqrt{2\,\hbar}}\,
\left\langle\,A\,e_l,\ \left(A^{-1}\right)^*\,B^*\ x\,\right\rangle\,\psi(x)
\\[3mm]\nonumber
&=&
\frac 1{\sqrt{2\,\hbar}}\,
\left\langle\,e_l,\ A^*\,\left(A^{-1}\right)^*\,B^*\ x\,\right\rangle\,\psi(x)
\\[3mm]\nonumber
&=&
\frac 1{\sqrt{2\,\hbar}}\,
\left\langle\,e_l,\ B^*\ x\,\right\rangle
\,\psi(x)
\\[3mm]\nonumber
&=&
\frac 1{\sqrt{2\,\hbar}}\,
\left\langle\,B\,e_l,\,x\,\right\rangle
\,\psi(x),
\end{eqnarray}
which is what we had to show.
\ep

\vskip 5mm
\noindent{\bf Proof of Theorem 3.1}:\quad
By an induction on $|k|$,
we prove there is a $|k|^{\mbox{\scriptsize th}}$ order
polynomial $p_k$ in $d$ variables with the following
properties:
\begin{enumerate}
\item[$\bullet$]\quad$\ds P_k(A,\,\hbar,\,x)\ =\
p_k(|A|^{-1}\,x/\sqrt{\hbar})$.
\item[$\bullet$]\quad$\ds \left(\frac{\partial\phantom{i}}{\partial z}\right)^k
G(x,\,z)\ =\ p_k(|A|^{-1}\,x/\sqrt{\hbar}-\overline{U_A}\,z)\,G(x,\,z)$.
\end{enumerate}
The result then follows by setting $z=0$.
We never compute the polynomial $p_k$ because it
may be complicated. We also use the notation
$p_k(A,\,\hbar,\,x)=p_k((|A|^{-1}\,x/\sqrt{\hbar})$.

We define $p_0=1$.
Below, we inductively compute
$\ds \left(\frac{\partial\phantom{i}}{\partial z}\right)^{k+e_l}G(x,\,z)$.
For our second condition above to hold, 
$p_{k+e_l}$ must be defined via the sum of
formulas (\ref{dog2}) and (\ref{dog3}). This uniquely
defines the polynomial $p_{k+e_l}$.

For $k=0$, the result is trivial since $P_0(A,\,\hbar,\,x)=1$.

For the induction step, it is sufficient to do the following for an arbitrary
positive integer $l\le d$:\\
{\em Assuming we have already proved these for some $k$,
we prove them for the multi--index $k+e_l$.}

To do this, we begin by noting that
$$
\varphi_k(A,\,B,\,\hbar,\,0,\,0,\,x)\ =\
\frac 1{\sqrt{k!}}\
{\cal R}^k(\varphi_0(A,\,B,\,\hbar,\,0,\,0))(x).
$$
Also,
$$
\varphi_k(A,\,B,\,\hbar,\,0,\,0,\,x)\ =\
2^{-|k|/2}\ (k!)^{-1/2}\ P_k(A,\,\hbar,\,x)\,
\varphi_0(A,\,B,\,\hbar,\,0,\,0,\,x).
$$
So,
$$
{\cal R}^k(\varphi_0(A,\,B,\,\hbar,\,0,\,0))(x)\ =\
2^{-|k|/2}\ P_k(A,\,\hbar,\,x)\
\varphi_0(A,\,B,\,\hbar,\,0,\,0,\,x).
$$
Thus, when we apply the $l^{\mbox{\scriptsize th}}$
raising operator, the polynomial $P_k(A,\,\hbar,\,x)$
gets changed to $\ds\frac 1{\sqrt{2}}\ P_{k+e_l}(A,\,\hbar,\,x)$.

Assuming the induction hypothesis, when we differentiate
$\ds\frac{\partial^kG}{\partial z^k}$ with respect to $z_l$,
the $z_l$ derivative can act on the $G(x,\,z)$ or it can act on the
$p_k(|A|^{-1}\,x/\sqrt{\hbar}-\overline{U_A}\,z)$.
When it acts on the $G(x,\,z)$, we obtain
\be\label{dog2}
2\,
\left\langle\,U_A\,e_l,\
\left(|A|^{-1}\,x/\sqrt{\hbar}-\overline{U_A}\,z\right)\,\right\rangle\
p_k(|A|^{-1}\,x/\sqrt{\hbar}-\overline{U_A}\,z)\ G(x,\,z).
\ee
From the induction hypotheses,
this is a polynomial of degree $|k|+1$ evaluated at the argument
$|A|^{-1}\,x/\sqrt{\hbar}-\overline{U_A}\,z$.
Note that this result depends on the following calculation, with $G(x,\,z)$
written with the polar decomposition of $A$:

\bea\nonumber
\frac{\partial G}{\partial z_l}(x,\,z)&=&
\left(-\ \langle\,U_A\,e_l,\,\overline{U_A}\,z\,\rangle\,-\,
\langle\,U_A\,\overline{z},\,\overline{U_A}\,e_l\,\rangle\,+\,
\frac 2{\sqrt{\hbar}}\,\langle\,U_A\,e_l,\,|A|^{-1}\,x\,\rangle\right)\ G(x,\,z)
\\[3mm]\nonumber
&=&
2\,\left\langle\,U_A\,e_l,\
\left(|A|^{-1}\,x/\sqrt{\hbar}-\overline{U_A}\,z\right)\,\right\rangle\ G(x,\,z).
\eea

When the~ $\ds\frac{\partial\phantom{i}}{\partial z_l}$~
acts on the polynomial, we either get zero or a polynomial of degree
$|k|-1$.
\bea\nonumber
&&-\ \left\langle\,
\overline{(\nabla p_k)(|A|^{-1}\,x/\sqrt{\hbar}-\overline{U_A}\,z)},\
\overline{U_A}\,e_l\,\right\rangle\ G(x,\,z)
\\[3mm]\label{dog3}
&=&
-\ \left\langle\,U_A\,e_l,\
(\nabla p_k)(|A|^{-1}\,x/\sqrt{\hbar}-\overline{U_A}\,z)\,\right\rangle\
G(x,\,z).
\eea

Recall that
$$
(R_l\,\psi)(x)\ =\ -\ \sqrt{\frac\hbar 2}\
\frac 1{\,\overline{\varphi_0(A,\,B,\,\hbar,\,0,\,0,\,x)}\,}\
\left\langle\,A\,e_l,\, \nabla\,
\left(\,\overline{\varphi_0(A,\,B,\,\hbar,\,0,\,0,\,x)}\
\psi(x)\right)\,\right\rangle,
$$
and that from our induction hypothesis,
\bea\nonumber
&&\overline{\varphi_0(A,\,B,\,\hbar,\,0,\,0,\,x)}\
\varphi_k(A,\,B,\,\hbar,\,0,\,0,\,x)
\\[3mm]\nonumber
&=&
\pi^{-d/2}\,\hbar^{-d/2}\,
2^{-|k|/2}\,(k!)^{-1/2}\,|\det\,A|^{-1}\,p_k(A,\,\hbar,\,x)\,
\exp\left\{-\,\frac{\left\langle\,x,\,|A|^{-2}\,x\,\right\rangle}{\hbar}\right\}.
\eea
When computing ${\cal R}_l\,\varphi_k$ by this formula,
the gradient in ${\cal R}_l$ can act on
the exponential or the $p_k(A,\,\hbar,\,x)$.
When it acts on the exponential, we get
\bea\nonumber
&&2^{-|k|/2}\ (k!)^{-1/2}\ p_k(A,\,\hbar,\,x)\
\sqrt{\frac 2\hbar}\,\left\langle\,A\,e_l,\,|A|^{-2}\,x\,\right\rangle\,
\varphi_0(A,\,B,\,\hbar,\,0,\,0,\,x)
\\[3mm]\nonumber
&=&
2^{-(|k|+1)/2}\ \sqrt{k_l+1}\ ((k+e_l)!)^{-1/2}
\\[3mm]\label{dog4}
&&\hspace{2cm}\times\quad
2\,\left\langle\,U_A\,e_l,\,|A|^{-1}\,x/\sqrt{\hbar}\,\right\rangle
\,p_k(A,\,\hbar,\,x)\
\varphi_0(A,\,B,\,\hbar,\,0,\,0,\,x).
\eea

When the gradient in ${\cal R}_l$ acts on the $p_k(A,\,\hbar,\,x)$,
in ${\cal R}_l\,\varphi_k$, we get the term
\bea\nonumber
&&\hspace{-5mm}
-\ \sqrt{\frac\hbar 2}\ 2^{-|k|/2}\ (k!)^{-1/2}\,
\left\langle\,A\,e_l,\,\nabla_x(p_k(A,\,\hbar,\,x))\,\right\rangle\,
\varphi_0(A,\,B,\,\hbar,\,0,\,0,\,x)
\\[3mm]\nonumber
&\hspace{-1cm}=&\hspace{-5mm}
-\ 2^{-(|k|+1)/2}\,(k!)^{-1/2}\,
\left\langle A\,e_l,\ \sum_{j=1}^d\,
\langle\,e_j,\,(\nabla p_k)(A,\,\hbar,\,x)\,\rangle\,
|A|^{-1}\,e_j\right\rangle\,
\varphi_0(A,\,B,\,\hbar,\,0,\,0,\,x)
\\[3mm]\nonumber
&\hspace{-1cm}=&\hspace{-5mm}
-\ 2^{-(|k|+1)/2}\ (k!)^{-1/2}\,
\left\langle\,A\,e_l,\ |A|^{-1}\,(\nabla p_k)(A,\,\hbar,\,x)\,\right\rangle\,
\varphi_0(A,\,B,\,\hbar,\,0,\,0,\,x)
\\[3mm]\nonumber
&\hspace{-1cm}=&\hspace{-5mm}
-\ 2^{-(|k|+1)/2}\ \sqrt{k_l+1}\ ((k+e_l)!)^{-1/2}
\\[3mm]\label{dog5}
&&\hspace{3cm}\times\quad\left\langle\,U_A\,e_l,\
(\nabla p_k)(A,\,\hbar,\,x)\,\right\rangle\,
\varphi_0(A,\,B,\,\hbar,\,0,\,0,\,x).
\eea

From (\ref{dog2}) and (\ref{dog3}) with $z=0$,
we obtain
$$
2\ \left\langle\,U_A\,e_l,\ |A|^{-1}\,x/\sqrt{\hbar}\,\right\rangle\
p_k(A,\,\hbar,\,x)\ -\
\left\langle\,U_A\,e_l,\ (\nabla p_k)(|A|^{-1}\,x/\sqrt{\hbar})\,\right\rangle.
$$

From (\ref{dog4}) and (\ref{dog5}) and taking into account the
factor of~ $\sqrt{k_l+1}$~ in\\
${\cal R}_l(\varphi_k)=\sqrt{k_l+1}\,\varphi_{k+e_l}$,~ we obtain
\bea\nonumber
&&P_{k+e_l}(A,\,\hbar,\,x)
\\[3mm]\nonumber
&=&
2\ \left\langle\,U_A\,e_l,\ |A|^{-1}\,x/\sqrt{\hbar}\,\right\rangle\,
p_k(A,\,\hbar,\,x)\ -\
\left\langle\,U_A\,e_l,\ (\nabla p_k)(|A|^{-1}\,x/\sqrt{\hbar})\,\right\rangle.
\eea

The quantities of interest contain the same polynomial evaluated at the
appropriate arguments, and
$P_{k+e_l}(A,\,\hbar,\,x)=p_{k+e_l}(A,\,\hbar,\,x)$.
Since $l$ is arbitrary, with $1\le l\le d$, the result is true for all
multi-indices with order $|k|+1$, and the induction can proceed.\ep

\section{The Rodrigues Formula}
As an application of the Generating Function formula, we prove a
multi--dimensional Rodrigues formula for these polynomials.
The result is
\begin{thm}\label{Rodrigues}
In $d$--dimensions, with the convention that $|A|$ is the positive square
root of $A\,A^*$,
$$
P_k(A,\,\hbar,\,x)\ =\ \exp\left(\frac{\| |A|^{-1}x\|^2}{\hbar}\right)\
\left(-\,\sqrt{\hbar}\,A^*\nabla_x\right)^k\
\exp\left(-\,\frac{\| |A|^{-1}x\|^2}{\hbar}\right).
$$
\end{thm}

\vskip 5mm
\noindent
{\bf Remark}~
By scaling, it is sufficient to prove this for $\hbar=1$.

\vskip 5mm
We begin the proof of this result with a lemma that embodies
a special case of the chain rule
for high order derivatives in $d$--dimensions.
\begin{lem}
Assume $F:{\mathbb R}^d\to{\mathbb R}^d$ is $C^m$ and that
$M:{\mathbb R}^d\to{\mathbb R}^d$ is linear. Then
viewing gradients as column vectors and using multi--index notation,
$$
\left(\nabla_x\right)^m\ F(M\,x)\ =\
\left(\left(\,M^t\ \nabla_y\right)^m\,F\right)_{y=M\,x}
$$
\end{lem}
One proves this by induction on $|m|$, proving that each component is correct.

\noindent
{\bf Proof of Theorem 4.1}\quad Using this technical result
and the generating function,
\begin{eqnarray}\nonumber
\hspace{-13mm}&&P_k(A,\,1,\,x)
\\[3mm]\nonumber
\hspace{-13mm}&=&\left(\nabla_z\right)^k
\exp\left(-\,(\overline{U_A}\,z)^t\,(\overline{U_A}\,z)\,+\,
2\,(\overline{U_A}\,z)^t\,(|A|^{-1}\,x)\right)_{z=0}
\\[3mm]\nonumber
\hspace{-13mm}&=&\left(\nabla_z\right)^k
\exp\left(-\,(\overline{U_A}\,z\,-\,|A|^{-1}\,x)^t\,
(\overline{U_A}\,z\,-\,|A|^{-1}\,x)\right)\
\exp\left((|A|^{-1}\,x)^t\,(|A|^{-1}\,x)\right)_{z=0}
\\[3mm]\nonumber
\hspace{-13mm}&=&\exp\left((|A|^{-1}\,x)^t\,(|A|^{-1}\,x)\right)\
\left(\nabla_z\right)^k
\exp\left(-\,(\overline{U_A}\,z\,-\,|A|^{-1}\,x)^t\,
(\overline{U_A}\,z\,-\,|A|^{-1}\,x)\right)_{z=0}
\\[3mm]\nonumber
\hspace{-13mm}&=&\exp\left((|A|^{-1}\,x)^t\,(|A|^{-1}\,x)\right)\,
\left((\overline{U_A})^t\,\nabla_w\right)^k
\exp\left(-\,(w\,-\,|A|^{-1}\,x)^t\,
(w\,-\,|A|^{-1}\,x)\right)_{w=0}
\\[3mm]\nonumber
\hspace{-13mm}&=&\exp\left((|A|^{-1}\,x)^t\,(|A|^{-1}\,x)\right)\,
\left((\overline{U_A})^t\,\nabla_w\right)^k
\exp\left(-\,(w\,-\,u)^t\,
(w\,-\,u)\right)_{\hspace{-4mm}
\scriptsize\begin{array}{c}w=0\\ u=|A|^{-1}x\end{array}}
\\[3mm]\nonumber
\hspace{-13mm}&=&\exp\left((|A|^{-1}\,x)^t\,(|A|^{-1}\,x)\right)\,
\left(-\,(\overline{U_A})^t\,\nabla_u\right)^k
\exp\left(-\,(w\,-\,u)^t\,
(w\,-\,u)\right)_{\hspace{-4mm}
\scriptsize\begin{array}{c}w=0\\ u=|A|^{-1}x\end{array}}
\\[3mm]\nonumber
\hspace{-13mm}&=&\exp\left((|A|^{-1}\,x)^t\,(|A|^{-1}\,x)\right)\,
\left(-\,(\overline{U_A})^t\,|A|^t\,
\nabla_x\right)^k
\exp\left(-\,(w\,-\,|A|^{-1}\,x)^t\,
(w\,-\,|A|^{-1}\,x)\right)_{w=0}
\\[3mm]\nonumber
\hspace{-13mm}&=&\exp\left((|A|^{-1}\,x)^t\,(|A|^{-1}\,x)\right)\,
\left(-\,(\overline{U_A})^t\,|A|^t\,
\nabla_x\right)^k
\exp\left(-\,(|A|^{-1}\,x)^t\,(|A|^{-1}\,x)\right).
\end{eqnarray}
However, $|A|^t=|A|$ is real,
and with our unusual convention, $A=|A|\,U_A$.
So, $(\overline{U_A})^t\,|A|=A^*$.\\
This proves the theorem.\ep


\begin{thebibliography}{xxxxxxx}
\bibitem{DKT}Dietert, H., Keller, J., and Troppmann, S.:\quad An Invariant
Class of Hermite Type Multivariate Polynomials for the Wigner Transform.
(2015 preprint,  arXiv:1505.06192).

\bibitem{raise}Hagedorn, G.A.:\quad Raising and Lowering Operators for
Semiclassical Wave Packets.
{\it Ann.~Phys.} {\bf 269}, 77--104 (1998).

\bibitem{Keller}Keller, J.F.:\quad Quantum Dynamics on Potential Energy
Surfaces.
{\it Doctoral Thesis}, Technische Universit\"at -- M\"unchen (2015).

\bibitem{O}Ohsawa, T.:\quad private communication (2015).
\end{thebibliography}
\end{document}